\documentclass{article}

\usepackage{graphicx}%
\usepackage{multirow}%
\usepackage{amsmath,amssymb,amsfonts}%
\usepackage{amsthm}%
\usepackage{mathrsfs}%
\usepackage[title]{appendix}%
\usepackage{xcolor}%
\usepackage{textcomp}%
\usepackage{manyfoot}%
\usepackage{booktabs}%
\usepackage{algorithm}%
\usepackage{algorithmicx}%
\usepackage{algpseudocode}%
\usepackage{listings}%
\usepackage[a4paper,margin=2cm]{geometry}
\usepackage{hyperref}

\theoremstyle{plain}%
\theoremstyle{definition}%

\theoremstyle{remark}%

\newenvironment{dedication}
  {\clearpage           
   \thispagestyle{empty}
   \vspace*{\stretch{1}}
   \raggedleft          
   \section*{Dedication} 
   \itshape             

  }
  {\par 
   \vspace{\stretch{3}} 
   \clearpage           
  }

\begin{document}

\title{A Levinson's theorem for particle form factors}

\author{%
Francesco Rosini\thanks{francesco.rosini@pg.infn.it. These authors contributed equally to this work.} \\
Simone Pacetti\thanks{simone.pacetti@unipg.it. These authors contributed equally to this work.} \\
\small Dipartimento di Fisica e Geologia, Universita degli Studi di Perugia, Via Alessandro Pascoli, Perugia 06123, Italy \\
\small Instituto Nazionale di Fisica Nucleare, Via Alessandro Pascoli, Perugia 06123, Italy
}
\date{}

\maketitle


\begin{abstract}
We present and demonstrate a version of Levinson's theorem especially dedicated to the asymptotic behavior of form factor phases. Indeed, as required by analyticity, form factors are multi-valued complex functions of a square four-momentum defined in the complex plane with a cut along the positive real axis. Their phases evaluated on the upper edge of this cut, i.e., on the time-like region, tend asymptomatically to integer multiples of $\pi$ radians. The Levinson's theorem establishes a univocal relation between such multiples and properties of form factors related to the dynamics of the electromagnetic interaction of the corresponding hadrons.
\end{abstract}


%




\begin{dedication}
To the memory of our friend and colleague Rinaldo Baldini Ferroli, who passed away on 21 February 2026. From those who knew him as a friend, and from one who regrets not having had the opportunity to meet him but who continues to view the world through the legacy of his work, we dedicate this paper to him as a sign of gratitude for his contributions to the field and for his passion for physics.
\end{dedication}
\section{Introduction}\label{sec:introduction}
Analyticity of form factors (FFs) is widely exploited in particle phenomenology to extract valuable information on the structure of hadrons and on the dynamics of their electromagnetic interactions (see Refs.~\cite{PhysRevD.110.054003, PhysRevD.104.116016, hammer2006nucleon,HAMMER1996343} and related citations). In this manuscript we will present and prove a version of the Levinson's theorem~\cite{titchmarsh1968theory}, which classically relates the asymptotic limit of the phase of scattering amplitudes and the number of bound states, to the case of hadrons' FFs.
\\
In quantum field theory, FFs are irreducible Lorentz scalar functions appearing in the most general description of matrix elements of the electromagnetic current $J^\mu$ between hadronic states~\cite{Lepage:1979za}. The set of FFs of a hadron $h$ $\{F_j(s)\}_j$ is defined through the expression
\begin{equation}
\langle h(p',\lambda')|J^\mu(0)|h(p,\lambda)\rangle=\bar{u}(p',\lambda')\left[\sum_{j}F_j(s)C^\mu_j\right]u(p,\lambda)\,,
\end{equation}
where $p$ ($p'$) is the initial (final) four-momentum, $\lambda$ ($\lambda'$) is the initial (final) polarization, $s=q^2=(p'-p)^2$ is the squared four-momentum transfer, $\{C^\mu_j\}_j$ is the set of covariant structures allowed by Lorentz invariance. They have to transform collectively as a vector under parity transformation and can be expressed in term of the elements of the Clifford algebra $\mathrm{Cl}_{1,3}(\mathbb{R})$~\cite{clifford1871}
\begin{equation}
  \{\mathbb{I},\gamma^\mu,\sigma^{\mu\nu},\gamma^\mu\gamma_5,\gamma_5\}\,,
\end{equation}
where $\gamma^\mu$ are the Dirac matrices, $\sigma^{\mu\nu}=i[\gamma^\mu,\gamma^\nu]/2$, $\gamma_5=i\gamma^0\gamma^1\gamma^2\gamma^3$ and $\mathbb{I}$ is the identity element of the algebra. The number of independent FFs is $2S+1$, where $S$ is the spin of the hadron. For instance, spin-$1/2$ baryons have two independent FFs, a possible choice is that of the pair of so-called Dirac and Pauli FFs. 
\\
The FFs are analytic functions of the single variable $s$ on $s$-complex plane with the cut $[s_{\rm th},\infty)$ along the positive real axis. The branch point $s_{\rm th}$ is called theoretical threshold and it is the squared mass of the lightest hadronic state that can couple to the $h\bar h$ system. If such a system has an isovector component, the theoretical threshold is $s_{\mathrm{th}}=(2m_\pi)^2$, corresponding to the $\pi^+\pi^-$ hadronic state. For purely isoscalar $h\bar h$ states, the theoretical threshold is instead $s_{\rm th}=(2m_\pi+m_{\pi^0})^2$, because the lightest hadronic state is that of three pions $\pi^+\pi^-\pi^0$.
\\
The analyticity of FFs allows to relate their properties in the space-like region ($s<0$), where they are real functions, to those in the time-like region ($s>0$), where are complex functions, with, in general, non-vanishing imaginary parts. Data on FFs of a hadron $h$ can be extracted from cross sections of reactions whose Feynman diagrams contain the vertex $h\bar h\gamma$. Space-like real values of FFs are obtained from the cross section of the lepton-hadron ($l$-$h$) scattering process $lh\to lh$, time-like moduli can be extracted by studying the differential cross section of the hadron-anti-hadron production via lepton-anti-lepton annihilation process, $l^+l^-\to h\bar h$.
\\
The FFs obey the Phragm\'en-Lindel\"of theorem~\cite{titchmarsh1968theory}, which, for a complex function $F(z)$ analytic and bounded in the slice region of the complex plane delimited by the straight lines $\{z:z=Re^{i\theta_{1}},R\in\mathbb{R}\}$ and $\{z:z=Re^{i\theta_{2}},R\in\mathbb{R}\}$, states that
\begin{equation}
\lim_{R\to \infty}F(Re^{i\theta_1})=
\lim_{R\to \infty}F(Re^{i\theta_2})\,,
\label{eq:PL}
\end{equation}
with $\theta_1,\theta_2\in[0,2\pi)$.\\ 
Assuming that the FF $F(s)$ is also bounded, being by definition analytic in complex plane with the cut on the positive real axis, the identity of Eq.~\eqref{eq:PL}, with $\theta_1=0^+$ and $\theta_2=\pi$ implies that the space-like and the time-like limits are the same, i.e.,
\begin{equation}
\underbrace{\lim_{s\to-\infty} F(s)}_{\rm space-like}
=	
\underbrace{\lim_{s\to\infty} F(s)}_{\rm time-like}\,.
\nonumber
\end{equation}
This also means that, since FFs are real in the space-like region, they must be so asymptotically in the time-like region, or, in other terms, their imaginary parts vanish faster then their real parts as the four-momentum squared $s$ diverges for positive values.
\section{Levinson's theorem}\label{sec:phragmen}
Consider a function $F(z)$ analytic in the domain $C=\left\{z:z\not\in[x_0,\infty)\cup\{p_j\}_{j=1}^\nu\right\}$, which is the complex plane deprived of the positive real cut $[x_0,\infty)\subset\mathbb{R}$ and of the set $\{p_j\}_{j=1}^\nu$ of $\nu$ isolated poles, whose corresponding orders are the elements of the set $\{n_j\}_{j=1}^\nu\subset\mathbb{N}$ and having also a set $\{z_k\}_{k=1}^\mu$ of $\mu$ isolated zeros with the set of orders $\{m_k\}_{k=1}^\mu\subset\mathbb{N}$. 
The function $F(z)$ can be factorized as
\begin{equation}
\label{eq:factorization}
    F(z)=f(z)\prod_{k=1}^\mu(z-z_k)^{m_k}\prod_{j=1}^\nu\frac{1}{(z-p_j)^{n_j}}\,,
\end{equation}
where the function $f(z)$ has neither zeroes nor poles in $\mathbb{C}$, and, as the original $F(z)$, is a multi-valued function having the same cut $[x_0,\infty)$ on the positive real axis.
\begin{figure}[h]
\centering
\includegraphics[width=0.5\textwidth]{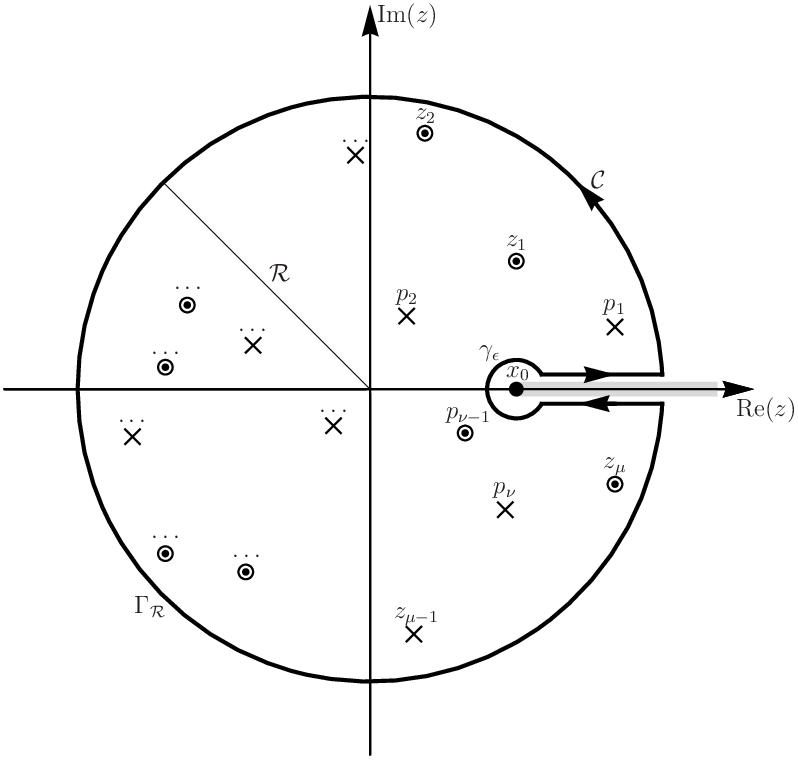}
\caption{Integration path $\mathcal{C}$ for the integral of Eq.~\eqref{eq:levinson}. The gray band on the positive real axis indicates the cut. Rounded dots and crosses represent zeros and poles, respectively.}
\label{fig:integrationpath}
\end{figure}\\
Using the argument principle~\cite{beee1599-a4ec-3a3b-bd94-6636e5310dcc} on the closed contour ${\cal C}$, shown in Fig.~\ref{fig:integrationpath}, which is the union of two straight segments and two arcs, $\Gamma_{\cal R}$ and $\gamma_\epsilon$, centered in the origin and on the real threshold and branch point $x_0$, with radii ${\cal R}$ and $\epsilon$, respectively.
\\
Assuming: $\mathcal{R}>\max_{(k,j)\in\{1,\ldots,\mu\}\times\{1,\ldots,\nu\}}\{|z_k|,|p_j|\}$, which implies that the winding number of the curve ${\cal C}$ around each pole and each zero is one, we have
\begin{equation}
\label{eq:levinson}
    \frac{1}{2i\pi}\oint_{\mathcal{C}}\frac{d\ln [F(z)]}{dz}dz=M-N\,,
\end{equation}
where
\begin{equation}
	M=\sum_{k=1}^\mu m_k\,,\ \ \ 	N=\sum_{j=1}^\nu n_j\nonumber
\end{equation}
are the total numbers of zeros and poles counted with their multiplicities. In order to obtain this result, the logarithmic derivative of $F(z)$ can be expressed using the factorization in Eq.~\eqref{eq:factorization} as
\begin{equation}
\label{eq:logderivative}
\frac{d\ln [F(z)]}{dz}=\frac{d\ln [f(z)]}{dz}+\sum_{k=1}^\mu\frac{m_k}{z-z_k}-\sum_{j=1}^\nu\frac{n_j}{z-p_j}\,,
\nonumber
\end{equation}
then the integral of Eq.~\eqref{eq:levinson} counts the residuals of the zeroes with a positive sign, while those of the poles are considered with a negative sign. Since the logarithmic derivative of the function $f(z)$ is analytic in the complex plane with the cut $[x_0,\infty)$, we have
\begin{equation}
\oint_{\cal C}\frac{d\ln[f(z)]}{dz}dz=0	\,,
\nonumber
\end{equation}
for each pair of radii $({\cal R},\epsilon)\in(0,\infty)^2$, as a consequence of the Cauchy's integral theorem. Moreover, assuming that the function $F(z)$ is real for real values of $z$, it holds the Schwarz reflection principle 
\begin{equation}
\label{eq:schwarz}
F(z^*)=F^*(z)\,.
\end{equation}
The first consequence is that all the singularities  lying outside the real axis are pairs of complex conjugates. Secondly, the integral of Eq.~\eqref{eq:levinson}, in the limits ${\cal R}\to\infty$ and $\epsilon\to0^+$, can be written as the sum of four contributions, i.e.,
\begin{align}
\label{eq:levinson2}
\lim_{{\cal R}\to\infty}
\lim_{{\epsilon}\to0^+}
\frac{1}{2i\pi}\oint_{\mathcal{C}}\frac{d\ln [F(z)]}{dz}dz =& 
\frac{1}{2i\pi}\left[
\lim_{{\cal R}\to\infty}
\int_{\Gamma_\mathcal{R}}\frac{d\ln [F(z)]}{dz}dz\right.\\
&+
\lim_{{\epsilon}\to 0^+}
\left(
\int_{\gamma_\epsilon}\frac{d\ln [F(z)]}{dz}dz\right.\nonumber\\
&\left. \left.+\int_{x_0+i\epsilon}^{\infty +i\epsilon}\frac{d\ln [F(z)]}{dz}dz+\int_{\infty-i\epsilon}^{x_0-i\epsilon}\frac{d\ln [F(z)]}{dz}dz\right)\right]\,,\nonumber
\end{align}
where, as shown in Fig.~\ref{fig:integrationpath}, $\Gamma_\mathcal{R}$ is the arc of radius $\mathcal{R}$ centered in the origin, while $\gamma_\epsilon$ is the arc of radius $\epsilon$ centered in the branch point $x_0$. Due to the regularities of $F(z)$ in $z=x_0$, i.e.,
\begin{equation}
	F(z)\mathop{\sim}_{z\to x_0} Kz^\alpha\,,
	\nonumber
\end{equation}
where $K\in \mathbb{C}$ is a constant and ${\rm Re}(\alpha)>-1$, the integral over the path $\gamma_\epsilon$ is vanishing as $\epsilon\to 0^+$. Using this result and the Schwarz reflection principle of Eq.~\eqref{eq:schwarz}, the previous limit becomes
\begin{align}
\label{eq:levinson3}
\lim_{{\cal R}\to\infty}\lim_{{\epsilon}\to0^+}
\frac{1}{2i\pi}\oint_{\mathcal{C}}\frac{d\ln [F(z)]}{dz}dz
=&\frac{1}{\pi}\left(\lim_{{\cal R}\to\infty}\int_{{\Gamma}^+_{\mathcal{R}}}\mathrm{Im}\left[\frac{d\ln [F(z)]}{dz}\right]dz
\right.\nonumber\\
&+\left.
\int_{x_0}^{\infty}\mathrm{Im}\left[\frac{d\ln [F(x)]}{dx}\right]dx\right)\,,
\end{align}
where the arc ${\Gamma}^+_\mathcal{R}$ is the portion of the arc $\Gamma_\mathcal{R}$ lying in the half plane of positive imaginary parts, i.e.,
\begin{equation}
{\Gamma}^+_\mathcal{R}=\{z:z=\mathcal{R}e^{i\theta},\theta\in\left( 0, \pi \right]\}\,.
\nonumber
\end{equation}
Indicating with $\phi(z)$ the phase of the function $F(z)$, i.e., $\phi(z)=\arg[F(z)]$, we have
\begin{equation}
	{\rm Im}\{\ln[F(z)]\}=
	{\rm Im}\{\ln|F(z)|+i\arg[F(z)]\}=\arg[F(z)]=\phi(z)\,,
	\nonumber
\end{equation}
hence, the second integral in the left-hand side of Eq.~\eqref{eq:levinson3} gives
\begin{align}
\label{eq:levinson5}
\frac{1}{\pi}\int_{x_0}^{\infty}\mathrm{Im}\left[\frac{d\ln[F(x)]}{dx}dx\right]
&=
\frac{1}{\pi}\int_{x_0}^{\infty}\frac{d\mathrm{Im}\{\ln[F(x)]\}}{dx}dx\nonumber\\
&=
\frac{1}{\pi}\int_{x_0}^{\infty}\frac{d\phi(x)}{dx}dx
\nonumber\\
&=\frac{\phi(\infty)-\phi(x_0)}{\pi}\,.
\end{align}
To compute the first integral on the right-hand side of Eq.~\eqref{eq:levinson3}, we exploit the Cauchy-Riemann equations, which, for the real and imaginary parts of a function $g(z)$ analytic in a neighborhood of the point $z_0$, read
\begin{equation}
\label{eq:cauchy-riemann}
\left.\frac{dg}{dz}\right|_{z=z_0}=\left({\partial \mathrm{Re}[g]\over \partial x}+i{\partial \mathrm{Im}[g]\over \partial x}\right)_{x=x_0,y=y_0}=\left({\partial \mathrm{Im}[g]\over \partial y}-i{\partial \mathrm{Re}[g]\over \partial y}\right)_{x=x_0,y=y_0},
\end{equation}
where $z=x+iy$ and $z_0=x_0+iy_0$. For values of the radius ${\cal R}$, such that  $\mathcal{R}>\max_{(k,j)\in\{1,\ldots,\mu\}\times\{1,\ldots,\nu\}}\{|z_k|,|p_j|\}$, the function $F(z)$ has neither singularities nor poles on ${\Gamma}^+_\mathcal{R}$, so that the function $\ln[F(z)]$ is analytic and, using the first identity of Eq.~\eqref{eq:cauchy-riemann},
\begin{align}
\label{eq:levinson4}
    &\frac{1}{\pi}\int_{{\Gamma}^+_\mathcal{R}}\mathrm{Im}\left[\frac{d\ln[F(z)]}{dz}\right]dz=
    \frac{1}{\pi}\int_{\Gamma^+_\mathcal{R}}\mathrm{Im}\left[{{\partial \ln |F(z)|}\over {\partial x}}+i{{\partial \phi(z)}\over {\partial x}}\right]dz =\frac{1}{\pi}\int_{\Gamma^+_\mathcal{R}}{{\partial \phi(z)}\over {\partial x}}dz\,.
\end{align}
To compute this integral in the limit ${\cal R}\to\infty$, we use the Phragm\'en-Lindel\"of theorem, that over the real axis gives
\begin{equation}
\label{eq:asymptoticphase}
    \phi(x)\mathop{\sim}_{x\to\infty} \phi_{\infty}+x^{-a} \,,
\end{equation}
with $a\in(0,\infty)$, as $x\to \infty$, and where $\phi_{\infty}$ is a real constant. Indeed, since the space-like and time-like limits coincide and since in the space-like region the function $F(z)$ is real, the phase must fulfill
\begin{equation}
    \lim_{x\to \infty}\phi(x)\equiv \phi_\infty=n\pi\,,
    \nonumber
\end{equation}
where $n\in\mathbb{N}$. This means that phase has an horizontal asymptote, as it follows from Eq.~\eqref{eq:asymptoticphase}. However, the Phragm\'en-Lindel\"of theorem, expressed in Eq.~\eqref{eq:PL}, extends this behavior in the whole upper half $z$-plane, so that
\begin{equation}
    \phi(z)\mathop{\sim}_{\mathcal{R}\to\infty}\phi_\infty+{\cal R}^{-a}\,,
    \label{eq:asy-z}
\end{equation}
with $z={\cal R}e^{i\theta}$ and $\theta \in\left(0,\pi\right]$. The asymptotic behavior of the phase in the upper-half complex plane can be given in the form of Eq.~\eqref{eq:asy-z} because, for large values of $|z|$, it does not depend of the phase $\theta$, so that it is a real function of $|z|={\cal R}$. The modulus of the last integral of Eq.~\eqref{eq:levinson4}, and hence the integral itself, goes to zero as the radius $\cal R$ diverges, indeed
\begin{align}
\label{eq:levinson6}
    \left|\int_{{\Gamma}^+_\mathcal{R}}\frac{\partial \phi(z)}{\partial x}dz \right|
\sim    
\left|\int_{{\Gamma}^+_\mathcal{R}}\frac{d}{d x}\left(
\phi_\infty+x^{-a}
\right)dz \right|
=\left|a\int_{{\Gamma}^+_\mathcal{R}}x^{-a-1}dz \right|
\le|a|\pi {\cal R}^{-a}\mathop{\to}_{{\cal R}\to\infty}0\,.
\nonumber
\end{align}
On the light of this result and of that of Eq.~\eqref{eq:levinson5}, the integral of Eq.~\eqref{eq:levinson3} is
\begin{equation}
\lim_{{\cal R}\to\infty}\lim_{{\epsilon}\to0^+}
\frac{1}{2i\pi}\oint_{\mathcal{C}}\frac{d\ln [F(z)]}{dz}dz	=\frac{\phi(\infty)-\phi(x_0)}{\pi}\,.
\nonumber
\end{equation}
On the other hand, even under the limits ${\cal R}\to\infty$ and $\epsilon\to0^+$, the argument principle of Eq.~\eqref{eq:levinson} is valid, i.e.,
\begin{equation}
\lim_{{\cal R}\to\infty}\lim_{{\epsilon}\to0^+}
\frac{1}{2i\pi}\oint_{\mathcal{C}}\frac{d\ln [F(z)]}{dz}dz	=M-N\,.
\nonumber
\end{equation}
Finally, using the previous two expression we obtain
\begin{equation}
    \phi(\infty)-\phi(x_0)= \pi(M-N)\,,
    \label{eq:levi0}
\end{equation}
which represents the version of Levinson's theorem we would like to propose. 
%
%
\section{Phase of a hadronic form factor}
It is interesting to study explicitly the case of a generic hadronic FF, because, from its power-law asymptotic behavior in the space-like region, inferred by exploiting the QCD constituent-quark counting rules~\cite{Matveev:1973ra,PhysRevLett-31-1153}, we can obtain the time-like asymptotic value of the phase.
\\
Consider a generic hadronic FF $G(s)$, analytic in the $s$ complex-plane deprived of the real, positive branch-cut $[s_{\rm th},\infty)$, where the branch-point $s_{\rm th}$ is the theoretical threshold (see Sec.~\ref{sec:introduction}). For real values of $s$, the FF is real if $s<s_{\rm th}$, while it has a non-vanishing imaginary part for $s>s_{\rm th}$. Using logarithmic dispersion relations~\cite{Geshkenbein:1969bb} (DRs), under the assumption of no zeros for the FF, the time-like phase of $G(t)$, i.e., for $t>s_{\rm th}$, is given by the integral in principal value
\begin{equation}
\label{eq:DR-phase}
    \delta(t)=-\frac{\sqrt{t-s_{\rm th}}}{\pi}\Pr\!\!\int_{s_{\rm th}}^{\infty}\frac{\ln|G(s)|}{\sqrt{s-s_{\rm th}}(s-t)}ds\,, \ \ \ \ t>s_{\rm th}\,,
\end{equation}
where $\delta(t)$ is defined through its relation with the FF: $G(t)=|G(t)|e^{i\delta(t)}$. The previous expression is also known as DR for the phase.
\\
Counting rules of perturbative QCD~\cite{Matveev:1973ra,PhysRevLett-31-1153} predict that, as $s$ diverges in the space-like region, i.e., $s\to-\infty$, hadronic FFs vanish as a negative, integer power of $s$. In the case under study, this implies that
\begin{equation}
G(s)={\cal O}\left( s^{-n}\right)\,,\ \ \ \mbox{as: }s\to-\infty\,,
\nonumber
\end{equation}
for a certain $n\in\mathbb{N}$.
Assuming that the FF $G(s)$ is bounded and analytic in the whole upper-half complex plane $s$, the one of positive imaginary parts, the same asymptotic behavior can be considered in the time-like region, as a consequence of the Phragm\'en-Lindel\"of theorem (see Sec.~\ref{sec:introduction}). In order to study the asymptotic region, we consider a FF already in its power-law form at the theoretical threshold, i.e.,
\begin{equation}
    G(s)= G_{\rm th}\left(\frac{s_{\rm th}-s_1}{s-s_1}\right)^n\,,\ \ \ \mbox{for: }s\ge s_{\rm th}\,,
    \label{eq:power-law}
\end{equation}
where $G_{\rm th}\in\mathbb{C}$ is the threshold value of the FF and $s_1>s_{\rm th}$. Note that, to satisfy the applicability conditions of the DRs, $s_1$ must lie outside the complex domain bounded by the closed contour ${\cal C}$ shown in Fig~\ref{fig:integrationpath}.
\\
Let us substitute the expression for $G(s)$ in Eq.~\eqref{eq:DR-phase}. Since the integral is evaluated between $s_{\rm th}$ and $t>s_{\rm th}$, we can omit the absolute value in the logarithm at the numerator. Using the usual properties of the logarithms we obtain
\begin{equation}
    \delta(t)=\frac{n}{\pi}\sqrt{t-s_{\rm th}}\Pr\!\!\int_{s_{\rm th}}^\infty\frac{\ln(s-s_1)-\ln(s_{\rm th}-s_1)-\ln|G_{\rm th}|/n}{\sqrt{s-s_{\rm th}}(s-t)}ds\,.
    \label{eq:drl-0}
\end{equation}
We perform the substitutions $s\to u$, with
\begin{equation}
e^u=\sqrt{\frac{s-s_{\rm th}}{t-s_{\rm th}}}\,, \ \ \  e^udu=\frac{ds}{2\sqrt{s-s_{\rm th}}\sqrt{t-s_{\rm th}}}\,, \ \ \  
s=(t-s_{\rm th})e^{2u}+s_{\rm th}\,,
\nonumber
\end{equation}
so that the integral of Eq.~\eqref{eq:drl-0} becomes
\begin{align}
    \delta(t)&=
    \frac{2n}{\pi}
    \Pr\!\!
    \int_{-\infty}^\infty\frac{\ln\left[(t-s_{\rm th})e^{2u}+s_{\rm th}-s_1\right]-\ln(s_{\rm th}-s_1)-\ln|G_{\rm th}|/n}{e^{2u}-1}e^u du
    \nonumber\\
    &=\frac{n}{\pi}
    \Pr\!\!
    \int_{-\infty}^\infty\frac{\ln\left[(t-s_{\rm th})e^{2u}+s_{\rm th}-s_1\right]-\ln(s_{\rm th}-s_1)-\ln|G_{\rm th}|/n}{\sinh(u)} du\nonumber\\
    &=\frac{n}{\pi}
    \Pr\!\!
    \int_{-\infty}^\infty\frac{\ln\left[(t-s_{\rm th})e^{2u}+s_{\rm th}-s_1\right]}{\sinh(u)} du\,,
    \nonumber
\end{align}
where the contribution proportional to $\left(\ln(s_{\rm th}-s_1)+\ln|G_{\rm th}|/n\right)$ is null since
\begin{equation}
	\Pr\!\!
    \int_{-\infty}^\infty\frac{du}{\sinh(u)}=0\,.
    \nonumber
\end{equation}
By using this result, we may also write
\begin{align}
    \delta(t)		
    &=\frac{n}{\pi}
    \Pr\!\!
    \int_{-\infty}^\infty\frac{\ln\left[e^{2u}+(s_{\rm th}-s_1)/(t-s_{\rm th})\right]+\ln(t-s_{\rm th})}{\sinh(u)} du
    \nonumber\\
    &=\frac{n}{\pi}
    \Pr\!\!
    \int_{-\infty}^\infty\frac{\ln\left[e^{2u}+(s_{\rm th}-s_1)/(t-s_{\rm th})\right]}{\sinh(u)} du\,,
    \nonumber
\end{align}
in this form, the limit $t\to\infty$ can be directly taken by neglecting the term $s_{\rm th}/(t-s_{\rm th})$. It follows that the asymptotic limit of the phase is
\begin{align}
 \lim_{t\to\infty}  \delta(t)		
    &=\lim_{t\to\infty}
    \frac{n}{\pi}
    \Pr\!\!
    \int_{-\infty}^\infty\frac{\ln\left[e^{2u}+(s_{\rm th}-s_1)/(t-s_{\rm th})\right]}{\sinh(u)} du
    =
    \frac{2n}{\pi}
    \Pr\!\!
    \int_{-\infty}^\infty\frac{u}{\sinh(u)} du
    \,,
\nonumber
\end{align}
and, being the last integral well known as
\begin{equation}
	\Pr\!\!
    \int_{-\infty}^\infty\frac{u}{\sinh(u)} du=\frac{\pi^2}{2}\,,
    \nonumber
\end{equation}
the limit value of the phase is
\begin{equation}
\lim_{t\to\infty}  \delta(t)=\delta(\infty)=n\pi\,.
\nonumber
\end{equation}
Moreover, since from the DR for the phase of Eq.~\eqref{eq:DR-phase},	we have $\delta(s_{\rm th})=0$, the previous result can be written in the form of the Levinson's identity of Eq.~\eqref{eq:levi0}, as
\begin{equation}
\delta(\infty)-\delta(s_{\rm th})=n\pi	\,.
\label{eq:weird}
\end{equation}
This appears contradictory, because the FF $G(s)$ is analytic, hence pole-free, and is also assumed to be zero-free in $s$ complex plane with the cut $[s_{\rm th},\infty)$. As a consequence we should find
\begin{equation}
\delta(\infty)-\delta(s_{\rm th})=0	\,.
\end{equation}
Where does the result of Eq.~\eqref{eq:weird} come from? How should this be interpreted?
\\
One possibility is the following: in order to reproduce the asymptotic behavior predicted by perturbative QCD, the FF must vanish asymptotically as a negative integer power of $s$, $s^{-n}$. This implies the emergence of a set of $n$ corresponding poles. In the example we treated, a single pole $s_1$ of order $n$ has been considered. This pole, lying outside the domain bounded by closed contour ${\cal C}$, is enclosed by such a curve in the clockwise direction, for all values of the radii ${\cal R}$ and $\epsilon$, and hence the integral of Eq.~\eqref{eq:levinson}, written for the FF $G(s)$, becomes
\begin{equation}
	\frac{1}{2i\pi}\oint_{\cal C}\frac{d\ln[G(s)]}{ds}ds=-(-n)=n\,,
\nonumber
\end{equation}
where the first minus sign arises from the fact that $s_1$ is a pole, and the second from the fact that it is enclosed in the negative (i.e., clockwise) direction. This result leads directly to that of Eq.~\eqref{eq:weird}.
\\
The power-law behavior at high squared momentum transfer in the space-like region represents the unveiling of perturbative QCD dynamics. It manifests itself in the amplitudes, and consequently in the FFs, as a product of gluon propagators, each one proportional to $s^{-1}$. These propagators represent the gluon exchanges which are required to share the momentum transferred by the virtual photon among the valence quarks of the hadron, ensuring it remains intact. We can assume that QCD condensation at low energies dresses the gluon propagators, providing them with finite masses. This moves the pole from the origin to the time-like region, above the theoretical threshold. Such a straightforward interpretation can explain the effective power-law defined in Eq.~\eqref{eq:power-law}.
\section{Conclusion}\label{sec13}
This work presents a version of the standard Levinson's theorem specifically tailored for hadronic FFs. It relates the time-like asymptotic limit of the phase to: the value of the phase itself at the theoretical threshold,  the number of zeros (given the analyticity of the FF, there are no poles) and the power of $s^{-1}$ of its asymptotic behavior. The expression follows directly from Eqs.~\eqref{eq:levi0} and \eqref{eq:weird}. Consider a FF $G(s)=|G(s)|e^{i\delta(s)}$, with the theoretical threshold $s_{\rm th}$, $M$ zeros, and the asymptotic power law $s^{-n}$, the resulting relationship reads
\begin{equation}
\delta(\infty)-\delta(s_{\rm th})=\pi(M+n)\,.
\nonumber
\end{equation}
This is our final result, which states that both the power-law, ruling the asymptotic behavior and the presence of zeros have the same enhancing effect on the limit of the phase. This also means that the only FF with the same phase value at the theoretical threshold and at infinity is the one that is zero-free and goes asymptotically to a constant different from zero.


\bibliographystyle{unsrt}
\bibliography{sn-bibliography}

\end{document}